\documentclass[prb,twocolumn,superscriptaddress,showpacs,
amsmath,amssymb]{revtex4}

\usepackage{graphicx} % Include figure files
\usepackage{dcolumn} % Align table columns on decimal point
\newcommand{\e}{\mathrm e}

\def\Gam{\Gamma}

\def\Lam{\Lambda}
\def\om{\omega}

\def\bq{{\bf q}}

\def\bx{{\bf x}}
\def\b0{{\bf 0}}

\def\cO{{\cal O}}

\def\tq{\tilde q}

\def\tT{\tilde T}
\def\tu{\tilde u}
\def\tdelta{\tilde\delta}

\begin{document}

\title{Critical temperature and Ginzburg region near a quantum critical
 point in two-dimensional metals}

\author{J.~Bauer}
\email{j.bauer@fkf.mpg.de}
\affiliation{Max-Planck-Institute for Solid State Research,
 D-70569 Stuttgart, Germany}
\author{P.~Jakubczyk}
\affiliation{Institute of Theoretical Physics, Faculty of Physics, Warsaw University, 
 Ho\.za 69, 00-681 Warsaw, Poland}
\author{W.~Metzner}
\affiliation{Max-Planck-Institute for Solid State Research,
 D-70569 Stuttgart, Germany}

\date{\today}

\begin{abstract}
We compute the transition temperature $T_c$ and the Ginzburg temperature
$T_{\rm G}$ above $T_c$ near a quantum critical point at the boundary of an 
ordered phase with a broken discrete symmetry in a two-dimensional 
metallic electron system.
Our calculation is based on a renormalization group analysis of the
Hertz action with a scalar order parameter.
We provide analytic expressions for $T_c$ and $T_{\rm G}$ as a function of the 
non-thermal control parameter for the quantum phase transition, 
including logarithmic corrections.
The Ginzburg regime between $T_c$ and $T_{\rm G}$ occupies a sizable part 
of the phase diagram.
\end{abstract}

\maketitle

%%% Introduction %%%%%%%%%%%%%%%%%%%%%%%%%%%%%%%%%%%%%%%%%%%%%%%%%%%%%%%%%%

\section{Introduction}

Instabilities of the normal metallic state lead to a rich variety of
quantum phase transitions in interacting electron systems.
Near a quantum critical point electronic excitations are strongly 
scattered by order parameter fluctuations such that Fermi liquid 
theory breaks down.\cite{vojta03,loehneysen07}
The fluctuation effects and the ensuing non-Fermi liquid behavior
is particularly pronounced in two-dimensional systems.
It is therefore not surprising that quantum critical fluctuations 
are frequently invoked as a mechanism for the enigmatic strange metal
behavior observed in cuprate superconductors and other layered 
correlated electron compounds.

Theoretical works have mostly focused on the quantum critical point and 
its extension into the quantum critical regime at finite temperature,
where quantum fluctuations are particularly important.
Less attention has been paid to the {\em Ginzburg region} near the
critical temperature, which is characterized by strongly interacting 
classical order parameter fluctuations.\cite{goldenfeld92}
This is somewhat unwarranted since classical critical fluctuations
also affect electronic excitations very strongly.
In two dimensions they lead to a contribution of the order $T\xi$ to 
the quasi-particle decay rate, where $\xi$ is the diverging correlation
length.\cite{vilk97,abanov03,katanin05,dellanna06}

In this paper we compute the size of the Ginzburg region above the
critical temperature near a quantum critical point in two-dimensional 
metals.
More specifically, we consider continuous quantum phase transitions 
associated with the spontaneous breaking of a {\em discrete} symmetry, 
described by an effective Hertz action \cite{hertz76} for a scalar order 
parameter with dynamical exponents $z=2$ or $z=3$.
We calculate the transition temperature $T_c$ as a function of the 
non-thermal control parameter for the quantum phase transition,
as well as the Ginzburg temperature $T_{\rm G}$ above $T_c$. The size of 
the Ginzburg region $T_{\rm G} - T_c$ is determined to leading order
in the distance from the quantum critical point.
The dependence of the Ginzburg temperature on the control parameter 
was derived already by Millis.\cite{millis93} 
However, that study did not access 
the Ginzburg region between $T_{\rm G}$ and $T_c$.
A comprehensive analysis of all finite temperature transition and
crossover lines in dimensions $d>2$ was performed by Sachdev.
\cite{sachdev97}
For discrete symmetry breaking in two dimensions the critical 
temperature $T_c$ and the Ginzburg temperature below $T_c$ were 
recently compared within a renormalization group study which allowed 
to approach the finite temperature transition.\cite{jakubczyk08}
However, in that work the flow equations were solved only numerically,
while we now present analytic results.

The paper is organized as follows. In Section II we derive the renormalization
group equations for the effective Hertz action. These are solved analytically 
in an approximate form in Section III. In Section IV we discuss the results 
for $T_c$, $T_{\rm G}$ and the size of the Ginzburg region, before concluding in
Section V.

%%% Hertz action %%%%%%%%%%%%%%%%%%%%%%%%%%%%%%%%%%%%%%%%%%%%%%%%%%%%

\section{Hertz action and flow equations}

Our analysis is based on the Hertz action \cite{hertz76}
\begin{eqnarray}
 {\cal S}[\phi] &=& 
 \frac{T}{2} \sum_{\om_n} \int \frac{d^dq}{(2\pi)^d}
 \phi_{\bq,\om_n} 
 \left( \delta_0 + \bq^2 + \frac{|\omega_{n}|}{|\bq|^{z-2}} \right) 
 \phi_{-\bq,\om_n} \nonumber\\ 
 &&+ 
 \frac{u_0}{4!} \int_0^{\frac{1}{T}} \! d\tau \int d^d x \, 
 \phi^4(\bx,\tau) \; ,
 \label{action}
\end{eqnarray}
where $\phi(\bx,\tau)$ is a real scalar order parameter field and 
$\phi_{\bq,\om_n}$ its momentum representation;
$\omega_n = 2\pi n T$ with integer $n$ denotes the bosonic 
Matsubara frequencies.
For the dynamical exponent $z$ we consider the cases $z=2$, which
describes density wave transitions, and $z=3$, relevant for a 
nematic transition or Ising-type ferromagnetic transitions.
We do not address the issue under which circumstances the Hertz
action provides a faithful description of quantum criticality in
two-dimensional metals.\cite{nonHertz}

Before embarking on the renormalization group approach, we would
like to emphasize that in two dimensions $T_c(\delta_0)$ cannot be 
obtained from a first order expansion in the quartic coupling $u_0$, 
even if it is weak and irrelevant at the quantum critical point.
To leading order in $u_0$, the inverse susceptibility $\delta$ is
given by
\begin{equation}
 \delta = \delta_0 + a \, T \sum_{\om_n} \int \frac{d^dq}{(2\pi)^d} \,
 \frac{u_0}{\delta_0 + \bq^2 + \frac{|\om_n|}{|\bq|^{z-2}}} \; ,
\end{equation}
where $a$ is a positive constant.
At finite temperature the Matsubara frequencies are discrete and
the classical fluctuation contribution from $\om_n = 0$ diverges
logarithmically in the limit $\delta_0 \to 0$ in two dimensions.
Trying to treat this divergence by a self-consistent equation,
replacing $\delta_0$ by $\delta$ under the integral, one finds that 
the transition temperature $T_c$ is suppressed to zero at the 
critical point given by $\delta = 0$, irrespective of $\delta_0$.
This behavior is reminiscent of the Mermin-Wagner theorem, which 
excludes spontaneous breaking of a continuous symmetry in two 
dimensions. 
However, the above first order calculation is essentially 
independent of the symmetry of the order parameter, and is 
therefore misleading at least in the case of a discrete symmetry.

We solve the problem by using flow equations which describe
the renormalization of the inverse susceptibility (or ''mass'') 
$\delta$ and the quartic coupling $u$ due to fluctuations.
The flow equations are derived from an approximate ansatz for the
exact effective action $\Gam^{\Lam}[\phi]$, that is, the generating
functional for vertex functions in the presence of an infrared
cutoff $\Lam$.\cite{berges02} 
The cutoff is implemented by adding a regulator term of the form 
$\frac{1}{2} \int \phi R^{\Lam} \phi$ to the bare action 
${\cal S}[\phi]$.
The exact flow of $\Gam^{\Lam}[\phi]$ is given by the Wetterich 
equation \cite{wetterich93}
\begin{equation}
 \partial_{\Lam} \Gam^{\Lam}[\phi] = \frac{1}{2} {\rm tr} 
 \frac{\partial_{\Lam} R^{\Lam}}{\Gam^{(2)\Lam}[\phi] + R^{\Lam}} 
 \; ,
\label{flow_exact}
\end{equation}
where $\Gam^{(2)\Lam}[\phi]$ is the matrix of second derivatives
of $\Gam^{\Lam}[\phi]$ with respect to $\phi$ and the trace sums
over momenta and frequencies.
We approximate $\Gam^{\Lam}[\phi]$ by an ansatz of the form
Eq.~(\ref{action}) with a renormalized mass term $\delta^{\Lam}$ 
and a renormalized coupling $u^{\Lam}$. Inserting this ansatz in 
the exact flow equation for $\Gam^{\Lam}[\phi]$ and comparing
coefficients, one obtains \cite{remark1}
\begin{equation}
  \partial_{\Lam} \delta^{\Lam} = 
 - \frac{1}{2} u^{\Lam} T \sum_{\om_n}\!\! \int\!\! \frac{d^d q}{(2\pi)^d}\!\!\!
 \, \frac{\partial_{\Lam} R^{\Lam}(\bq)}
 {\left[ \delta^{\Lam} + \bq^2 + \frac{|\om_n|}{|\bq|^{z-2}}
 + R^{\Lam}(\bq) \right]^2} \; ,
 \label{flow_delta} 
\end{equation}
\begin{equation}
   \partial_{\Lam} u^{\Lam} = 
 3 (u^{\Lam})^2 T\!\! \int\!\! \frac{d^d q}{(2\pi)^d}\!\!\!
 \, \frac{\partial_{\Lam} R^{\Lam}(\bq)}
 {\left[ \delta^{\Lam} + \bq^2 + \frac{|\om_n|}{|\bq|^{z-2}}
 + R^{\Lam}(\bq) \right]^3} \; .
 \label{flow_u}
\end{equation}
The initial conditions for the flow are $\delta^{\Lam_0} = \delta_0$ 
and $u^{\Lam_0} = u_0$, where $\Lam_0$ is a (fixed) ultraviolet
cutoff.
As a regulator we choose the Litim \cite{litim01} function
$R^{\Lam}(\bq) = (\Lam^2 - \bq^2) \Theta(\Lam^2 - \bq^2)$, 
with derivative 
$\partial_{\Lam} R^{\Lam}(\bq) = 2\Lam \Theta(\Lam^2 - \bq^2)$,
which restricts the momentum integrals in the flow equations to 
$|\bq| \leq \Lam$ and replaces the $\bq^2$-term in the denominators 
by $\Lam^2$.
From now on we fix the dimensionality to $d$$=$$2$.

The Matsubara sums in the above flow equations can be expressed 
in terms of polygamma functions $\Psi_n(z)$, defined as
the $n$-th derivative of the digamma function 
$\Psi_0(z) = \Gam'(z)/\Gam(z)$.
Explicit $\Lam$-dependencies can be removed from the right hand side
of the flow equations as usual by introducing rescaled dimensionless
variables
\begin{equation}
 \tq = \frac{|\bq|}{\Lam} , \quad
 \tT = \frac{2\pi T}{\Lam^z} , \quad
 \tdelta = \frac{\delta^{\Lam}}{\Lam^2} , \quad
 \tu = \frac{T}{2\pi\Lam^2} u^{\Lam} \; .
\label{tildevar}
\end{equation}
One then obtains
\begin{equation}
 \Lam \partial_{\Lam} \tdelta =
 - 2 \tdelta - \frac{\tu}{2(1 + \tdelta)^2} -
 \frac{2\tu}{\tT^2} \int_0^1 d\tq \, \tq^{2z-3} \,
 \Psi_1[h(\tdelta,\tq,\tT)] \; , 
\label{flow_tdelta}  
\end{equation}
\begin{equation}
   \Lam \partial_{\Lam} \tu =
 - 2 \tu + \frac{3\tu^2}{(1 + \tdelta)^3} -
 \frac{6\tu^2}{\tT^3} \int_0^1 d\tq \, \tq^{3z-5} \,
 \Psi_2[h(\tdelta,\tq,\tT)] \; ,
\label{flow_tu}
\end{equation}
where $h(\tdelta,\tq,\tT)=1 + (1+\tdelta)\tq^{z-2}/\tT $.
The first term in each equation is due to the factor 
$\Lam^{-2}$ in the definition of the dimensionless variables, 
the second one captures classical fluctuations ($\om_n = 0$), 
and the third one quantum fluctuations ($\om_n \neq 0$).

%%%%%%%%%%%%%%%%%%%%%%%%%%%%%%%%%%%%%%%%%%%%%%%%%%%%%%%%%%%%%%%%%%%%%%%%%%%%%%%%%%

\section{Solution of flow equations}

For sufficiently small but finite temperature the flow
passes through two distinct regimes, which are distinguished 
by the size of the rescaled temperature $\tT$. 
Initially one has $\tT \ll 1$, such that quantum fluctuations
dominate, while in the final stage, for $\tT \gg 1$, the flow
is governed by classical fluctuations. In the latter regime
the third term on the right hand side of the flow equations
(\ref{flow_tdelta}) and (\ref{flow_tu}) can be neglected.
For $\tT \ll 1$ one can use the expansion of the polygamma
functions for large arguments, $\Psi_1(z) \approx z^{-1}$ and
$\Psi_2(z) \approx - z^{-2}$, to approximate the integrals in
Eqs.~(\ref{flow_tdelta}) and (\ref{flow_tu}) as
\begin{eqnarray}
 \frac{1}{\tT^2} \!\!\int_0^1 \!\!d\tq \, \tq^{2z-3} \,
 \Psi_1[ h(\tdelta,\tq,\tT)] &\approx&\!\!\!
 {\phantom -} \frac{1}{z(1+\tdelta) \tT} \; ,
 \label{int_tdelta} \\
 \frac{1}{\tT^3} \!\!\int_0^1 \!\!d\tq \, \tq^{3z-5} \,
 \Psi_2[h(\tdelta,\tq,\tT)] &\approx&\!\!\!
  \frac{-1}{z(1+\tdelta)^2 \tT} \; .
 \label{int_tu}
\end{eqnarray}
The cutoff scale corresponding to $\tT = 1$ is given by
$\Lam_1 = (2\pi T)^{1/z}$.
Following Millis \cite{millis93} we approximate the flow by its 
quantum contribution with the expansion Eqs.~(\ref{int_tdelta}) 
and (\ref{int_tu}) for $\Lam > \Lam_1$, and we discard the 
quantum terms for $\Lam < \Lam_1$.

The flow equations (\ref{flow_tdelta}) and (\ref{flow_tu}) exhibit
a fixed point at $\tdelta^* = - \frac{1}{7}$, 
$\tu^* = \frac{144}{343}$, and $\tT^* = \infty$.
This fixed point is approached by the flow at the finite temperature
phase transition, and it describes (approximately) the classical 
non-Gaussian critical fluctuations at $T_c$.
There $|\tdelta|$ is small during the entire flow. 
Above $T_c$ it remains small until the fluctuation contributions 
to the flow saturate at small $\Lam$.
We therefore approximate $1 + \tdelta \approx 1$ in the denominators
of the flow equations, which allows us to solve them analytically.
Note that within this approximation fluctuation contributions are
symmetric under $\tdelta \mapsto - \tdelta$, while in the exact 
flow they are larger for $\tdelta < 0$ compared to $\tdelta > 0$, which leads
to a suppression of $T_c$.

In the quantum regime ($\Lam > \Lam_1$) the approximate flow 
equations have the form
\begin{eqnarray}
 \Lam \partial_{\Lam} \tdelta &=&
 - 2 \tdelta - \frac{2}{z \tT} \tu \; , \\
 \Lam \partial_{\Lam} \tu &=& 
 - 2 \tu + \frac{6}{z \tT} \tu^2 \; .
\end{eqnarray}
Recall that $\tT = 2\pi T/\Lam^z$ is also a flowing quantity.
The explicit solution for $z$$=$$3$ reads
\begin{eqnarray}
 \Lam^2 \tu^{\Lam} &=& \frac{\pi T}{C_3 - \Lam}
 \; , \\
 \Lam^2 \tdelta^{\Lam} &=& C'_3 + f(C_3,\Lam) \; ,
\end{eqnarray}
where 
\begin{equation}
f(x,\Lam) = \frac{1}{6} \Lam^2 + \frac{1}{3} x \Lam + 
 \frac{1}{3} x^2 \ln\left( 1 - \frac{\Lam}{x} \right) \; .  
\end{equation}
The integration constants $C_3$ and $C'_3$ are determined by the
initial conditions at $\Lam_0$ as
$C_3 = \pi T \Lam_0^{-2} \tu_0^{-1} + \Lam_0$ and
$C'_3 = \delta_0 - f(C_3,\Lam_0)$.
The solution for $z=2$ is given by
\begin{eqnarray}
 \Lam^2 \tu^{\Lam} &=& \frac{\frac23 \pi T}{C_2 - \ln\Lam}
 \; , \\
 \Lam^2 \tdelta^{\Lam} &=& C'_2 +
 \frac{1}{3} e^{2 C_2} 
 {\rm Ei}\left( 2\ln\Lam - 2 C_2 \right) \; ,
\end{eqnarray}
where ${\rm Ei(x)} = \int_{-\infty}^x dt \, t^{-1} e^t$.
The integration constants $C_2$ and $C'_2$ are determined 
by the initial conditions:
$C_2 = \frac23 \pi T \Lam_0^{-2} \tu_0^{-1} + \ln\Lam_0$ and
$C'_2 = \delta_0 - \frac{1}{3} e^{2 C_2} 
 {\rm Ei}\left( 2\ln\Lam_0 - 2 C_2 \right)$.

In the classical regime ($\Lam < \Lam_1$) the approximate flow
equations read
\begin{eqnarray}
 \Lam \partial_{\Lam} \tdelta &=&
 - 2 \tdelta - \frac{1}{2} \tu \; , \label{tdelta_flowapprox} \\
 \Lam \partial_{\Lam} \tu &=& 
 - 2 \tu + 3 \tu^2 \; . \label{tu_flowapprox}
\end{eqnarray}
The explicit solution has the form
\begin{eqnarray}
 \tu^{\Lam} &=& \frac{1}{C \Lam^2 + \frac{3}{2}} \; , 
 \label{tu_c} \\
 \Lam^2 \tdelta^{\Lam} &=& C' -
 \frac{1}{4C} \ln \left( \Lam^2 + \frac{3}{2C} \right) \; .
 \label{tdelta_c}
\end{eqnarray}
The integration constants $C$ and $C'$ are determined by the
boundary conditions $\tu^{\Lam_1} = \tu_1$ and $\tdelta^{\Lam_1} =
\tdelta_1$ at the scale $\Lam_1$, yielding 
$C = \Lam_1^{-2} \left( \tu_1^{-1} - \frac{3}{2} \right)$ and
$C' = \Lam_1^2 \tdelta_1 +  
\frac{1}{4C} \ln \left( \Lam_1^2 + \frac{3}{2C} \right)$. 

At $T$$=$$0$ one has $\Lam_1$$=$$0$ and the flow can be obtained by 
taking the zero temperature limit of the solution in the quantum
regime. 
For $z=3$ the unscaled quartic coupling $u^{\Lam}$ saturates at
the finite value
\begin{equation}
u'_0 = \frac{u_0 }{1 + \frac{1}{2\pi^2} u_0 \Lam_0}=
 \frac{2\pi^2}{C_3}  \; ,  
\end{equation}
for $\Lam \to 0$.
Note that the rescaled variable $\tu$ vanishes at $T=0$.
For a generic choice of $\delta_0$ the inverse susceptibility 
$\delta^{\Lam}$ scales to a finite value near $\delta_0$. 
At the quantum critical point,
\begin{equation}
\delta_0 = \delta_0^{\rm qc} = f\big( \frac{2\pi^2}{u'_0},\Lam_0 \big)
 = - \frac{\Lam_0^3}{18\pi^2} u'_0 + \cO({u'_0}^2)\; ,
\end{equation}
the inverse susceptibility scales to zero for $\Lam \to 0$.

For $z=2$ the quartic coupling $u^{\Lam}$ vanishes logarithmically
for $\Lam \to 0$. The inverse susceptibility remains generically
finite, except at the quantum critical point given by
\begin{equation}
\delta_0^{\rm qc} = \frac{\Lambda_0^2}{3}\exp\left(\frac{8\pi^2}{3u_0}\right){\rm
  Ei}\left(-\frac{8\pi^2}{3u_0}\right) \; .  
\end{equation}

%%%%%%%%%%%%%%%%%%%%%%%%%%%%%%%%%%%%%%%%%%%%%%%%%%%%%%%%%%%%%%%%%%%%%%%%%%%%%%%%%%

\section{Results for $T_c$ and $T_{\rm G}$}

The phase transition line in the $(\delta_0,T)$ phase diagram is
determined by the condition $\delta^{\Lam} \to 0$ for $\Lam \to 0$.
Using the solution for $\delta^{\Lam}$ in the classical regime,
Eq.~(\ref{tdelta_c}), this yields a condition on the integration 
constants $C$ and $C'$, namely $C' = \frac{1}{4C} \ln\frac{3}{2C}$.
The constants $C$ and $C'$ can be expressed in terms of the bare
variables $\delta_0$ and $u_0$ by matching the initial condition
for the classical flow at $\Lam_1$ to the solution of the flow in
the quantum regime.

For $z=3$, one obtains
\begin{equation}
 \delta_0^c - \delta_0^{\rm qc} =  - f\left( \frac{2\pi^2}{u'_0},(2\pi T)^{\frac 13}
 \right) - \frac{\ln\left[ 1 + \frac{2}{3} (2\pi T)^{\frac 23} C(T) \right]}{4C(T)} 
  \; ,
 \label{Tc_z3}
\end{equation}
where $C(T) = \frac{2\pi}{u'_0 T} - \frac{7}{2(2\pi T)^{2/3}}$.
Expanding for small temperatures $T$ yields
\begin{equation}
 \delta_0^c - \delta_0^{\rm qc} = 
 - \frac{u'_0}{24\pi} T \ln \frac{T_0}{T} + \frac{u'_0}{9\pi} T + 
 \cO\left( T^{4/3} \ln T \right) \; ,
\label{Tc_z3_exp}
\end{equation}
with $T_0 = \frac{8(2\pi)^5}{27 {u'_0}^3}$.
Note that dependencies on the ultraviolet cutoff $\Lam_0$ are
absorbed in $u'_0$ and $\delta_0^{\rm qc}$.
Inverting Eq.~(\ref{Tc_z3_exp}) to leading order in $T$ yields
\begin{equation}
  T_c(\delta_0) = \frac{24\pi}{u'_0} \frac{\delta_0^{\rm qc} - \delta_0}
 {\ln\big( \frac{A_0}{\delta_0^{\rm qc} - \delta_0} \big)} \; ,
\end{equation}
with $A_0 = \frac{u'_0}{24\pi} T_0$.

For $z=2$, we find
\begin{equation}
 \delta_0^c - \delta_0^{\rm qc} =-\frac{1}{3}\e^{2C_2}  
 {\rm Ei}\left( \ln(2\pi T)- 2C_2 \right) - 
 \frac{\ln\Big[ 1 + \frac{4\pi T}{3} C(T) \Big]}{4 C(T)},
\label{Tc_z2}
\end{equation}
where $C(T) = \frac{3}{4\pi T}\ln(T_1/T)$ with 
$T_1 = \e^{2C_2}/(2\pi\e) = 
 (2\pi\e)^{-1} \Lam_0^2 \exp\left(\frac{8\pi^2}{3u_0}\right)$.
For low temperatures this becomes 
\begin{equation}
 \delta_0^c - \delta_0^{\rm qc}\!\! =\!\!-\frac{\pi}{3} \frac{T
 \ln[\ln(T_1/T)]}{\ln(T_1/T)} + \frac{2\pi}{3}\frac{T}{\ln(T_1/T)} + 
 \cO\left( T/ (\ln T)^2 \right).
\label{Tc_z2_exp}
\end{equation}
Note that the dependence on $u_0$ enters only via the scale $T_1$.
Inverting Eq.~(\ref{Tc_z2_exp}) for small $T$ yields
\begin{equation}
T_c(\delta_0) = \frac{3}{\pi} \frac{(\delta_0^{\rm qc} - \delta_0) \ln\big(\frac{A_1}{\delta_0^{\rm qc} - \delta_0}\big)}
{\ln\big[ \ln\big( \frac{A_1}{\delta_0^{\rm qc} - \delta_0} \big) \big]} \; ,  
\end{equation}
with $A_1 = \frac{\pi}{3} T_1$.

The functional form of the leading temperature dependencies in 
Eqs.~(\ref{Tc_z3_exp}) and (\ref{Tc_z2_exp}) is consistent with a
numerical solution of the full flow equations (\ref{flow_tdelta})
and (\ref{flow_tu}). 
However, the prefactors are not reproduced exactly, since they are
affected by the approximations required for the sake of an analytic
solution. 
In Fig.~1 we show results for the critical line as obtained from
Eq.~(\ref{Tc_z3}) for $z=3$ and Eq.~(\ref{Tc_z2}) for $z=2$ for specific
choice of parameters.
For $z=3$ the logarithmic correction can be clearly seen, whereas for $z=2$ the
(partially compensating) logarithmic corrections are hardly visible, such that
the critical line looks almost linear.
\begin{figure}[!thbp]
\centering
\includegraphics[width=0.45\textwidth]{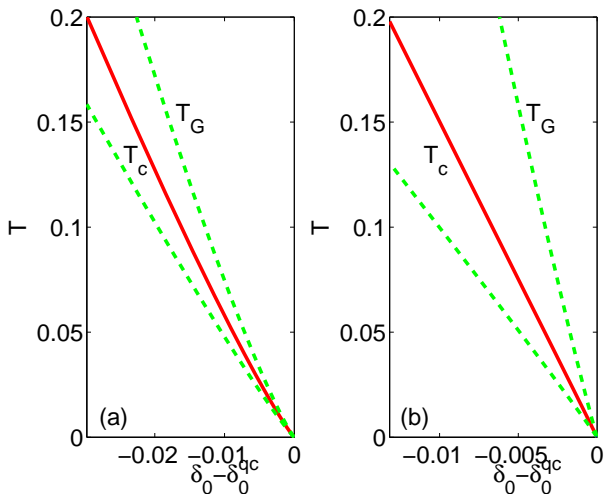}
\caption{(Color online): 
  Critical temperature $T_c$ and Ginzburg temperature $T_{\rm G}$ as a
  function of $\delta_0 - \delta_0^{\rm qc}$ for $z=3$ (a) and
  $z=2$ (b). 
  The bare $\phi^4$-coupling is $u_0 = 2$, and the ultraviolet
  cutoff $\Lam_0 = 1$ in all cases.
  The parameter specifying the Ginzburg criterion has been 
  chosen as $x=2/3$.
  The Ginzburg temperature below $T_c$ has not been computed 
  here, but has been added schematically for completeness.}
\label{fig1}
\end{figure}

We now compute the Ginzburg line in the phase diagram, which marks the
boundary of the non-Gaussian classical fluctuation regime above $T_c$.
Scaling in the critical regime is governed by the interacting 
Wilson-Fisher fixed point $(\tdelta^*,\tu^*)$. 
We determine the Ginzburg line from the condition  
$\tu(\Lam_{\rm G}) = x \tu^*$ with $0<x<1$, where $\Lam_{\rm G}$ is the scale 
at which $\tdelta$ crosses zero, that is, $\tdelta(\Lambda_{\rm G}) = 0$.
The parameter $x$ is a measure for the closeness of $\tu$ to the 
fixed point.
This criterion comes about as follows. 
Close to criticality $\tdelta$ is negative at the beginning of the 
flow, but for $T> T_c$ it eventually increases and diverges with 
the trivial scaling dimension (see Eq.~(\ref{tdelta_flowapprox})). 
If $\tdelta$ crosses zero whilst $\tu$ is still small, $\tu$ has
little influence on the flow of $\tdelta$ also for 
$\Lam < \Lam_{\rm G}$ such that we are in the Gaussian regime with 
mean-field exponents.
On the other hand, if $\tu$ has reached a value close to its fixed 
point at $\Lam = \Lam_{\rm G}$, it affects the flow of $\delta$
substantially leading to non-Gaussian scaling.
There is no unique choice of $x$ quantifying the 
``closeness'' to the fixed point. 
This reflects the fact that the Ginzburg line marks a crossover
regime and not a sharp transition.

The leading low-$T$ behavior of $\delta_0^{\rm G} - \delta_0^{\rm qc}$ 
computed from the Ginzburg criterion described above turns out
to be the same as that for $\delta_0^c - \delta_0^{\rm qc}$,
with the same prefactor, irrespective of the choice of $x$.
However, differences appear in the first subleading term.
For $z=3$, one obtains
\begin{equation}
 \delta_0^{\rm G} - \delta_0^{c} = 
 - \frac{u_0'}{8\pi}\ln(x) T \; ,
\label{TG_z3}
\end{equation}
and for $z=2$,
\begin{equation}
 \delta_0^{\rm G} - \delta_0^{c} = 
 - \frac{\pi}{3}\ln(x) \frac{T}{\ln(T_1/T)} \, ,
\end{equation}
at low temperatures. We have used the fixed point value $\tu^*=2/3$ as deduced
from Eq. (\ref{tu_flowapprox}). 
Note that the terms on the right hand sides are positive.
Solving for $T_{\rm G} - T_c$ as a function of 
$\delta_0^{\rm qc} - \delta_0$, one finds that $(T_{\rm G} - T_c)/T_c$ 
is of order $|\ln(\delta_0^{\rm qc} - \delta_0)|^{-1}$ for $z=3$,
and of order $[\ln|\ln(\delta_0^{\rm qc} - \delta_0)|]^{-1}$ for 
$z=2$. Hence, the size of the Ginzburg region $T_{\rm G} - T_c$ is 
practically of the order $T_c$ near the quantum critical point. 
By contrast, in three dimensions it is of order $T_c^2$.\cite{millis93} 
The results for $T_G$ are plotted in Fig.~\ref{fig1} for the same choice of
parameters as above.  We can see a substantial Ginzburg regime opening
between $T_{\rm G}$ and $T_c$. 
The $T_c$- and $T_{\rm G}$-lines merge when the quantum critical point is approached, 
$\delta_0\to\delta_0^{\rm qc}$, since that critical point is Gaussian.

%%%%%%%%%%%%%%%%%%%%%%%%%%%%%%%%%%%%%%%%%%%%%%%%%%%%%%%%%%%%%%%%%%%%%%%%%%%%%%%%%%

\section{Conclusions}

We have derived analytic expressions for the transition
temperature $T_c$ and the Ginzburg temperature $T_{\rm G}$ above $T_c$ as
a function of the non-thermal control parameter $\delta_0$ near a 
quantum critical point with a scalar (Ising universality class) order 
parameter in a two dimensional metal.
The calculations are based on flow equations derived from a 
perturbative renormalization group for the Hertz model.
The renormalization of the quartic coupling is crucial to avoid 
an artificial suppression of $T_c$ to zero in two dimensions.
Both $T_c$ and $T_{\rm G}$ are essentially proportional to 
$\delta_0 - \delta_0^{\rm qc}$, with logarithmic corrections 
depending on the dynamical exponent $z$.
For $T_{\rm G}$ we confirm the results by Millis.\cite{millis93}
For $T_c$ we obtain the same logarithmic corrections as for
$T_{\rm G}$, in agreement with earlier evidence from a numerical solution 
of flow equations for the symmetry broken phase.\cite{jakubczyk08}
Nevertheless, the size of the Ginzburg region $T_{\rm G} - T_c$, which
has been calculated analytically for the first time in this paper,
is practically proportional to the distance to the the quantum critical
point, $\delta_0 - \delta_0^{\rm qc}$.
Hence, the Ginzburg region with its large non-Gaussian classical 
fluctuations covers a substantial part of the phase diagram near 
a continuous quantum phase transition in two dimensional metals.
Electronic excitations are strongly scattered by order parameter 
fluctuations in that region, which can lead to enhanced decay 
rates, pseudogaps, and other unconventional electronic properties.

\begin{acknowledgments}
We are grateful to So Takei and Hiroyuki Yamase for valuable 
discussions, and to Nils Hasselmann for useful comments and a critical reading
of the manuscript. This work was supported by the German Research Foundation  
through the research group FOR 723.
\end{acknowledgments}

%%%%%%%%%%%%%%%%%%%%%%%%%%%%%%%%%%%%%%%%%%%%%%%%%%%%%%%%%%%%%%%%%%%%


\begin{thebibliography}{99}

%\bibitem{sachdev99} S.~Sachdev,
% {\em Quantum Phase Transitions}
% (Cambridge University Press, Cambridge, UK, 1999).

%\bibitem{sondhi97} S. L. Sondhi et al., 
% Rev. Mod. Phys. {\bf 69}, 315 (1997).

\bibitem{vojta03} M.~Vojta,
 Rep. Prog. Phys. {\bf 66}, 2069 (2003).

\bibitem{loehneysen07} H.~v.~L\"ohneysen, A.~Rosch, M.~Vojta, and P.~W\"olfle,
 Rev. Mod. Phys. {\bf 79}, 1015 (2007).

\bibitem{goldenfeld92} For a discussion of the Ginzburg criterion,
 see, for example, N. Goldenfeld,
 {\em Lectures on phase transitions and the renormalization group}
 (Addison Wesley, Reading, 1992).

\bibitem{vilk97} Y.~M.~Vilk and A.-M.S.~Tremblay,
 J. Phys. I (France) {\bf 7}, 1309 (1997).

\bibitem{abanov03} A.~Abanov, A.~V.~Chubukov, and J.~Schmalian,
 Adv. Phys. {\bf 52}, 119 (2003).

\bibitem{katanin05} A.~A.~Katanin, A.P.~Kampf, and V.Yu.~Irkhin,
 Phys. Rev. B {\bf 71}, 085105 (2005);
 A.~A.~Katanin,
 Phys. Rev. B {\bf 72}, 035111 (2005).

\bibitem{dellanna06} L.~Dell'Anna and W.~Metzner,
 Phys. Rev. B {\bf 73}, 045127 (2006).

\bibitem{hertz76} J.~A.~Hertz,
 Phys. Rev. B {\bf 14}, 1165 (1976).

\bibitem{millis93} A.~J.~Millis,
 Phys. Rev. B {\bf 48}, 7183 (1993).

\bibitem{sachdev97} S.~Sachdev,
 Phys. Rev. B {\bf 55}, 142 (1997).

\bibitem{jakubczyk08} P.~Jakubczyk, P.~Strack, A.~A.~Katanin,
 and W.~Metzner,
 Phys. Rev. B {\bf 77}, 195120 (2008).

%\bibitem{belitz05} 
% D. Belitz, T.R. Kirkpatrick, and T. Vojta,
% Rev. Mod. Phys. {\bf 77}, 579 (2005).

\bibitem{nonHertz} 
 For a discussion of this point, see, for example,
 Ar.~Abanov and A.~V.~Chubukov, 
 Phys. Rev. Lett. {\bf 93}, 255702 (2004); 
 D. Belitz, T. ~R.~Kirkpatrick, and T. Vojta,
 Rev. Mod. Phys. {\bf 77}, 579 (2005);
 H.~v.~L\"ohneysen, A.~Rosch, M.~Vojta, and P.~W\"olfle,
 Rev. Mod. Phys. {\bf 79}, 1015 (2007); 
 M.~A.~Metlitski and S.~Sachdev, 
 Phys. Rev. B {\bf 82}, 075127 (2010), 
 {\em ibid} {\bf 82}, 075128 (2010). 

\bibitem{berges02} 
 J. Berges, N. Tetradis, and C. Wetterich,
 Phys. Rep. {\bf 363}, 223 (2002).

\bibitem{wetterich93} C. Wetterich,
 Phys. Lett. B {\bf 301}, 90 (1993).

\bibitem{remark1} The derivation is analogous to the
 derivation of the flow equations in the symmetry-broken
 state in Ref.~\onlinecite{jakubczyk08}, and simplified 
 due to the absence of $Z$-factors.

\bibitem{litim01} D.F. Litim,
 Phys. Rev. D {\bf 64}, 105007 (2001).

\end{thebibliography}
\end{document}